\documentclass[aps,showpacs,superscriptaddress,preprint]{revtex4}%
\usepackage{amsfonts}
\usepackage{amsmath}
\usepackage{amssymb}
\usepackage{graphicx}%
\begin{document}
\title{Orbital magnetization of the electron gas on a two-dimensional kagom\'{e}
lattice under a perpendicular magnetic field}
\author{Zhigang Wang}
\affiliation{LCP, Institute of Applied Physics and Computational Mathematics, P.O. Box
8009, Beijing 100088, People's Republic of China}
\author{Zi-Gang Yuan}
\affiliation{State Key Laboratory for Superlattices and Microstructures, Institute of
Semiconductors, Chinese Academy of Sciences, P. O. Box 912, Beijing 100083,
People's Republic of China}
\affiliation{College of Science, Beijing University of Chemical Technology, Beijing 100029,
People's Republic of China}
\author{Zhen-Guo Fu}
\affiliation{LCP, Institute of Applied Physics and Computational
Mathematics, P.O. Box 8009, Beijing 100088, People's Republic of
China} \affiliation{State Key Laboratory for Superlattices and
Microstructures, Institute of Semiconductors, Chinese Academy of
Sciences, P. O. Box 912, Beijing 100083, People's Republic of China}
\author{Shu-Shen Li}
\affiliation{State Key Laboratory for Superlattices and Microstructures, Institute of
Semiconductors, Chinese Academy of Sciences, P. O. Box 912, Beijing 100083,
People's Republic of China}
\author{Ping Zhang}
\thanks{Author to whom correspondence should be addressed. Electronic
address: zhang\_ping@iapcm.ac.cn} \affiliation{LCP, Institute of
Applied Physics and Computational Mathematics, P.O. Box 8009,
Beijing 100088, People's Republic of China} \affiliation{Center for
Applied Physics and Technology, Peking University, Beijing 100871,
People's Republic of China}


\begin{abstract}
The orbital magnetization of the electron gas on a two-dimensional
kagom\'{e} lattice under a perpendicular magnetic field is
theoretically investigated. The interplay between the lattice
geometry and magnetic field induce nontrivial $k$-space Chern
invariant in the magnetic Brillouin zone, which turns to result in
profound effects on the magnetization properties. We show that the
Berry-phase term in the magnetization gives a paramagnetic
contribution, while the conventional term brought about by the
magnetic response of the magnetic Bloch bands produces a diamagnetic
contribution. As a result, the superposition of these two components
gives rise to a delicate oscillatory structure in the magnetization
curve when varying the electron filling factor. The relationship
between this oscillatory behavior and the Hofstadter energy spectrum
is revealed by selectively discussing the magnetization and its two
components at the commensurate fluxes of $f$=$1/4$, $1/3$, and
$1/6$, respectively. In particular, we reveal as a typical example
the fractal structure in the magnetic oscillations by tuning the
commensurate flux around $f$=$1/4$. The finite-temperature effect on
the magnetization is also discussed.

\end{abstract}
\maketitle

\section{Introduction}

Over the past three decades, the orbital dynamics of two-dimensional
(2D) electrons coupled to a uniform perpendicular magnetic field has
been a subject of special interest in condensed matter physics.
Great attention has been paid to investigate the well-known
Hofstadter butterfly-like diagram, a remarkable intricate, detailed,
and self-similar feature for the subband spectrum of single-particle
eigenenergies, in different 2D structures such as square \cite{Hof},
triangular \cite{Claro}, honeycomb \cite{Ram}, and kagom\'{e}
lattices \cite{YXiao}, lateral superlattices \cite{Schu},
superstructures with flat bands \cite{Ando}, quasiperiodic tilings
\cite{Behrooz}, fractal networks \cite{Gordon}, and rhombus tilings
\cite{Vidal}. This magnetic-field-induced frustration in 2D
structures, caused by the competition between the periodic potential
imposed by the lattice structure and the length scale provided by
the magnetic field, is the headstream of rich and novel physical
phenomena in these 2D systems. Among them the orbital magnetization
is a very interesting one and needs to be paid special attention due
to its intrinsic relationship with the unique topological structure
brought about by the interplay between the special lattice geometry
and magnetic subbands. Unfortunately, to date, there are very scarce
studies \cite{Avron, Avron1} in literature that devote to
enlightening this revealing relationship.

Motivated by this observation, in this paper, we study the orbital
magnetization properties of the electron gas on a two-dimensional kagom\'{e}
lattice under a perpendicular magnetic field. Since our attention is solely on
the orbital character in magnetization brought about by the interplay between
the kagom\'{e} lattice structure and the magnetic field, thus unlike most of
previous work, the kagom\'{e} lattice considered in this paper is spin
nonmagnetic in itself, namely, no spin structure is exposed in the absence of
the external magnetic field. The nonmagnetic kagom\'{e} lattice structure has
been either fabricated by modern patterning techniques
\cite{Mohan2003,Hig2000} or observed in reconstructed semiconductor surfaces
\cite{Tong1985}. In the former case, remarkably, the electron filling factor
(namely, the Fermi energy) can be readily controlled by applying a gate
voltage \cite{Shi2001}. Our lattice model is free from the constraint imposed
on the $\mathbf{k}\mathtt{\cdot}\mathbf{p}$ approximation used in the
extensively studied GaAs two-dimensional electron gas (2DEG), in which the
$\mathbf{k}\mathtt{\cdot}\mathbf{p}$ Hamiltonian is only valid around the
$\Gamma$ point in the Brillouin zone (BZ). In contrast, our lattice model
allows for any electron filling, which result in various Fermi-surface
topologies in the magnetic BZ, which in turn, as will be shown below, produces
profound effects on the orbital magnetization properties and the related
transport phenomenon.

The orbital magnetization for Bloch electrons combines two terms \cite{Avron,
Xiao}, one is the conventional orbital magnetic moment to characterize the
magnetic response of the single-particle energy spectrum, and the other is a
Berry-phase correction. These two terms play different roles in metallic and
insulating regions \cite{Wang2007, Wang2}. By varying the Fermi energy, we
obtain the follows: (i) Both the magnitudes of these two terms increase
(decrease) with increasing (decreasing) the Fermi energy in metallic regions;
(ii) In insulating regions the conventional term keeps a constant unchanged
value (quantum step), while the Berry-phase linearly varies with the Fermi
energy with a slope proportional to the system's Hall conductance; (iii) A
general fractal structure may occur in the orbital magnetization curve (as a
function of the Fermi energy) by tuning the commensurate flux.

The rest of this paper is organized as follows. In Sec. II we give the
Hofstadter spectrum of the kagom\'{e} lattice by diagonalizing the
corresponding tight-binding Hamiltonian in an external magnetic field. In Sec.
III the orbital magnetization and its two components are systematically
discussed by varying various system parameters. A summary is given in Sec. IV.

\section{Kagom\'{e} lattice in a magnetic field}

\begin{figure}[ptb]
\begin{center}
\includegraphics[width=0.3\linewidth]{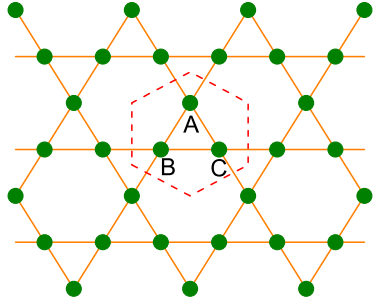}
\end{center}
\caption{(Color online). Schematic picture of the 2D kagom\'{e} lattice with
bond length $a$. The dashed line represents the Wigner-Seitz unit cell, which
contains three independent sites ($A$, $B$, and $C$).}%
\label{f1}%
\end{figure}

The kagom\'{e} lattice is a two-dimensional periodic array of corner-sharing
triangles with three sites per unit cell, as illustrated in Fig. \ref{f1}.
Here $a$ is the triangle edge length (we set it as the length unit in the
following), and $A$, $B$, and $C$ denote the three sites in a unit cell. Let
us begin with the tight-binding Hamiltonian for the spinless electrons on the
kagom\'{e} lattice with zero magnetic field ($\mathbf{B}$=$0$), which is given
by%
\begin{equation}
\mathcal{H}=t\sum_{\langle i,j\rangle}c_{i}^{\dag}c_{j}, \label{E1}%
\end{equation}
where $t$ is hopping amplitude between the nearest-neighbor link $\langle
i,j\rangle$ and $c_{i}^{\dag}$ ($c_{i}$) is the creation (annihilation)
operator of an electron on lattice $i$. Hamiltonian (\ref{E1}) can be
diagonalized in the momentum space as%
\begin{equation}
\mathcal{H}=t\sum_{\mathbf{k}}\Psi_{\mathbf{k}}^{\dag}H(\mathbf{k}%
)\Psi_{\mathbf{k}},
\end{equation}
where $\Psi_{\mathbf{k}}$=($c_{\mathbf{k}A}$, $c_{\mathbf{k}B}$,
$c_{\mathbf{k}C}$)$^{\text{T}}$ is the three-component electron field
operator. Each component of $\Psi_{\mathbf{k}}$ is the Fourier transform of
$c_{i}$, i.e.,%
\begin{equation}
c_{\mathbf{k}s}=\sum_{mn}c_{mns}e^{i\mathbf{k}\cdot\mathbf{r}_{mns}},
\end{equation}
where we have changed notation $i\rightarrow(mns)$ by using $(mn)$ to label
the kagom\'{e} unit cells and $s$ to label the three sites in a unit cell.
$H(\mathbf{k})$ is given by
\begin{equation}
H(\mathbf{k})=\left(
\begin{array}
[c]{ccc}%
0 & p_{\mathbf{k}}^{1} & p_{\mathbf{k}}^{3}\\
p_{\mathbf{k}}^{1} & 0 & p_{\mathbf{k}}^{2}\\
p_{\mathbf{k}}^{3} & p_{\mathbf{k}}^{2} & 0
\end{array}
\right)  ,
\end{equation}
where $p_{\mathbf{k}}^{i}$=$2\cos\left(  \mathbf{k}\cdot\mathbf{a}_{i}\right)
$ and $\mathbf{a}_{1}$=$(-1/2,-\sqrt{3}/2)$, $\mathbf{a}_{2}$=$(1,0)$, and
$\mathbf{a}_{3}$=$(-1/2,\sqrt{3}/2)$ represent the displacements in a unit
cell from the $A$ to $B$ site, from the $B$ to $C$ site, and from the $C$ to
$A$ site, respectively. In this notation, the first BZ is a hexagon with the
corners of $\mathbf{K}$=$\pm(2/3)\mathbf{a}_{1}$, $\pm(2/3)\mathbf{a}_{2}$,
and $\pm(2/3)\mathbf{a}_{3}$.

In the presence of a magnetic filed ($\mathbf{B}\neq0$), the hopping term of
the tight-binding Hamiltonian (\ref{E1}) is modified by phase factors from the
vector potential $\mathbf{A}$,%
\begin{equation}
t\rightarrow te^{i\gamma_{ij}}%
\end{equation}
where $\gamma_{ij}$ is the phase factor (the well known Peierls phase) between
sites $j$ and $i$:%
\begin{equation}
\gamma_{ij}=\frac{2\pi}{\Phi_{0}}\int_{j}^{i}\mathbf{A\cdot\mathrm{d}l}%
\end{equation}
with $\Phi_{0}$=$h/e$ being the flux quantum. 

One can take different gauge potential $\mathbf{A}$ mathematically to satisfy
the physical confinement $\nabla\times\mathbf{A}$\textbf{=}$\mathbf{B}$.
However, the flux through one unit cell is only linearly dependent on the
external magnetic field $\mathbf{B}$, and has no relation with the choice of
the gauge potential $\mathbf{A}$. So, it is convenient to measure the magnetic
field in units of the flux quantum per elementary triangular plaquette of the
kagom\'{e} lattice. After a simple algebraic calculation, one obtains the flux
through one triangle is $\Phi$=$\sqrt{3}B/4$. Thus we can define a parameter,
called the filling ratio $f$, as the fraction of a flux quantum throng each
triangle, i.e, $f$=$\Phi/\Phi_{0}$=$\sqrt{3}eB/4h$ \cite{YXiao}. The total
flux through a unit cell of the kagom\'{e} lattice is then given as
$8f\Phi_{0}$.

\begin{figure}[ptb]
\begin{center}
\includegraphics[width=0.5\linewidth]{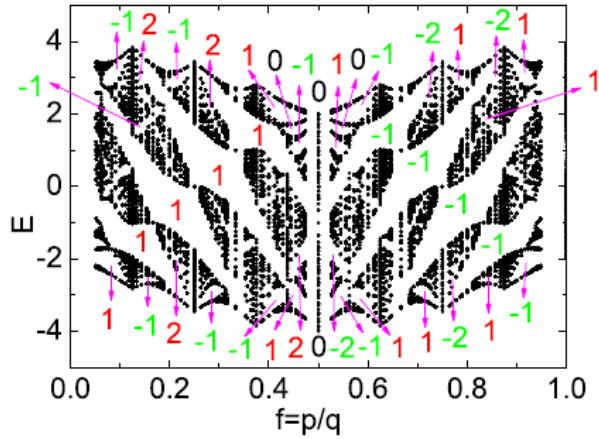}
\end{center}
\caption{(Color online). Hofstadter butterfly-like spectrum of the
2D Kagom\'{e} lattice. The numbers recorded in gaps represent the
Hall conductance of this system in units of $e^{2}/h$ when the Fermi
energy lies in these gaps. }%
\label{f2}%
\end{figure}

When $f$ is a rational number that can be written as $f$=$p/q$, where $p$ and
$q$ are integers with no common factors, the total flux through $q$ triangular
plaquette of the kagom\'{e} lattice is an integer multiple of $\Phi_{0}$. In
this case, the Hamiltonian (\ref{E1}) should be diagonalized in a
\textquotedblleft magnetic\textquotedblright\ unit cell, which contains $q$
plaquette in order to guarantee $\mathbf{k}$ being good quantum numbers. As a
result, the magnetic Brillouin zone (MBZ) is $q$ times smaller than the usual
BZ. Furthermore, because of the magnetic translation symmetry, the MBZ has
exactly a $q$-fold degeneracy. For convenience we take the gauge potential as
$\mathbf{A}$\textbf{=}$B(-\frac{1}{2}(-\sqrt{3}x+y),\frac{1}{2}(x-\frac
{1}{\sqrt{3}}y),0)$. Under this gauge, the Peierls phases, which reflect the
information about the external magnetic field, are only related to $n$ and
have nothing with $m$. The Hamiltonian of the kagom\'{e} lattice now can be
diagonalized in the momentum space as the following $3q\times3q$ matrix,

\begin{align}
&  H(\mathbf{k})=\sum_{n=0}^{q-1}\left[  h_{^{n}}^{BA}c_{nB}^{\dag}%
c_{nA}\right.  +h_{^{n}}^{BA\ast}c_{nB}^{\dag}c_{nA}\nonumber\\
&  +h_{n}^{CB}c_{nC}^{\dag}c_{nB}+\exp(-i4\pi f)h_{n+1}^{CB\ast}c_{nC}^{\dag
}c_{n+1B}\label{3q}\\
&  \left.  +h_{n}^{AC}c_{nA}^{\dag}c_{nC}+h_{^{n}}^{AC\ast}c_{n+1A}^{\dag
}c_{nC}+\text{H.c.}\right]  ,\nonumber
\end{align}
where
\begin{align*}
h_{n}^{BA}  &  =t\exp(-i\mathbf{k\cdot a}_{1})\exp\left[  -i8\pi f\left(
n-\frac{1}{6}\right)  \right]  ,\\
h_{n}^{CB}  &  =t\exp(-i\mathbf{k\cdot a}_{2})\exp\left[  i8\pi f\left(
n+\frac{1}{12}\right)  \right]  ,
\end{align*}
and
\[
h_{^{n}}^{AC}=t\exp(-i\mathbf{k\cdot a}_{3}).
\]
With the above Hamiltonian (\ref{3q}), it is straightforward to obtain the
Hofstadter butterfly-like energy spectrum of the kagom\'{e} lattice, which is
shown in Fig. \ref{f2}.

From Fig. \ref{f2}, one can easily observe three symmetries of this Hofstadter
butterfly-like spectrum of the kagom\'{e} lattice \cite{YXiao}: (i) The
spectrum at $f$ is the same as that at $f+j$ with $j$ any integer; (ii) The
spectrum is also unchanged on changing $f$ to $-f$, because if there is an
eigenstate with energy $\epsilon$ for field $f$, then the corresponding
complex conjugate of this eigenstate is also an eigenstate with the same
energy $\epsilon$ for field $-f$; (iii) The spectrum is inverted when the
field $f$ changes to $f+1/2$. Thus the highest-energy states for field near
$f$=$1/2$ are equivalent to the lowest-energy states near zero field.

\section{The properties of the orbital magnetization}

To obtain the orbital magnetization of the kagom\'{e} lattice in an external
magnetic field and with finite temperature, let us first write down the
single-particle free energy as follows \cite{Xiao}:
\begin{equation}
F=-\frac{1}{\beta}\sum_{n}\int_{\text{MBZ}}d^{2}\mathbf{k}\left[  1+\frac
{e}{\hslash}\mathbf{B}\cdot\Omega_{n}(\mathbf{k})\right]  \ln[1+e^{\beta
(\mu-\epsilon_{n\mathbf{k}})}].\label{free}%
\end{equation}
Here, $\Omega_{n}(\mathbf{k})$ is the Berry curvature of electronic Bloch
states defined by $\Omega_{n}(\mathbf{k})$=$i\langle\nabla_{\mathbf{k}%
}u_{n\mathbf{k}}|\times|\nabla_{\mathbf{k}}u_{n\mathbf{k}}\rangle$ with
$|u_{n\mathbf{k}}\rangle$ being the periodic part of Bloch wave for the $n$th
band. Its integral over the MBZ gives the topological invariant, namely, the
Chern number $C_{n}\mathtt{=-}\frac{1}{2\pi}\int_{\text{MBZ}}d^{2}%
\mathbf{k}\Omega_{n}(\mathbf{k})$. The sum of Chern numbers over the integer
occupied bands gives the quantized Hall conductance in unit of $e^{2}/h$ (see
numbers in Fig. 2). $\mu$ in Eq. (\ref{free}) is the electron chemical
potential, $\beta$=$1/k_{B}T$, and $\epsilon_{n\mathbf{k}}$ is the magnetic
band energy. When the field varies by an infinitesimal quantity, i.e., from
$\mathbf{B}$ to $\mathbf{B}\mathtt{+}\delta\mathbf{B}$, then the magnetic band
energy can be linearly expanded as $\epsilon_{n\mathbf{k}}(\mathbf{B}%
\mathtt{+}\delta\mathbf{B})$ = $\epsilon_{n\mathbf{k}}(\mathbf{B}%
)\mathtt{-}m_{n\mathbf{k}}(\mathbf{B})\cdot\delta\mathbf{B}$, where
$m_{n\mathbf{k}}$ is the conventional crystal orbital magnetic moment defined
by $m_{n\mathbf{k}}$=$-i(e/2\hslash)\langle\nabla_{\mathbf{k}}u_{n\mathbf{k}%
}|\times\left[  H(\mathbf{k})-\epsilon_{n\mathbf{k}}\right]  |\nabla
_{\mathbf{k}}u_{n\mathbf{k}}\rangle$ in the MBZ. The magnetization is then
given by the field derivative at fixed temperature and chemical potential,
$\mathcal{M}=-\left(  \partial F/\partial\mathbf{B}\right)  _{\mu,T}$, with
the result
\begin{align}
\mathcal{M} &  =\sum_{n}\int_{\text{MBZ}}d^{2}\mathbf{km}_{n}(\mathbf{k}%
)f_{n}(\mathbf{k})\label{m}\\
&  +\frac{1}{\beta}\sum_{n}\int_{\text{MBZ}}d^{2}\mathbf{k}\frac{e}{\hbar
}\mathbf{\Omega}_{n}\ln\left[  1+e^{\beta(\mu-\epsilon_{n\mathbf{k}})}\right]
\nonumber\\
&  \equiv\mathbf{M}_{c}+\mathbf{M}_{\mathbf{\Omega}},\nonumber
\end{align}
where $f_{n}(\mathbf{k})$ is the equilibrium Fermi-Dirac distribution function
for $n$th magnetic Bloch band. The first term in Eq. (\ref{m}) is just a
statistical sum of the orbital magnetic moments of the carriers originating
from the self-rotation of the carrier wave packet \cite{Chang,Sundaram}, thus
we call this term the conventional part $\mathbf{M}_{c}$ of the orbital
magnetization. Whereas the second term $\mathbf{M}_{\mathbf{\Omega}}$ is the
Berry-phase correction to the orbital magnetization. This term is of
topological nature. Interestingly, it is this Berry-phase term that eventually
enters the transport current \cite{Xiao2}. At zero temperature the general
expression (\ref{m}) reduces to \cite{Avron,Xiao,Thon1}%
\begin{equation}
\mathcal{M}=\sum_{n}\int_{\text{MBZ}}^{\mu_{0}}d^{2}\mathbf{k}\left[
\mathbf{m}_{n}(\mathbf{k})+\frac{e}{\hbar}\mathbf{\Omega}_{n}(\mathbf{k}%
)(\mu_{0}-\epsilon_{n\mathbf{k}})\right]  ,\label{m0}%
\end{equation}
where the upper limit means that the integral is over magnetic Bloch states
with energies below the Fermi energy $\mu_{0}$.

\begin{figure}[ptb]
\begin{center}
\includegraphics[width=0.4\linewidth]{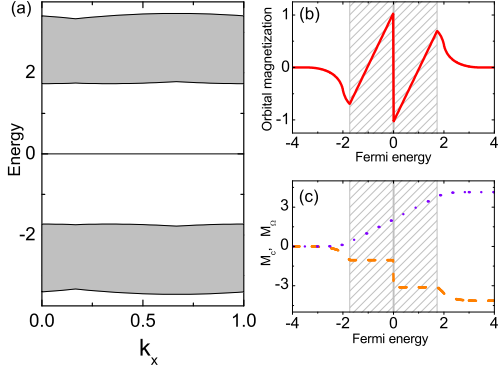}
\end{center}
\caption{(Color online). (a) Energy spectrum of the 2D kagom\'{e} lattice with
a field $f$=$1/4$. The shadow areas are the energy bands. The corresponding
magnetization $\mathcal{M}$ and its two components are respectively drawn in
(b) and (c) as functions of the Fermi energy $\mu_{0}$. The shaded areas in
(b) and (c) are the insulating regions. The dashed and dotted lines in (c)
represent $M_{c}$ and $M_{\Omega}$, respectively. }%
\label{f3}%
\end{figure}

Now with the help of the knowledge on the magnetic energy spectrum obtained
from diagonalizing the Hamiltonian (\ref{3q}), we can numerically obtain the
magnetization $\mathcal{M}$ for different field $f$. First, let us consider
the case with field $f$=$1/4$. In this case, the energy spectrum splits into
three bands between which there are two band gaps with the same gap width
$\Delta$=$1$.$74$ (see Fig. \ref{f3}(a)). The middle band collapse into a
plane in the MBZ. The calculated zero-temperature magnetization $\mathcal{M}$
is plotted in Fig. \ref{f3}(b). From this figure, one can find that with the
Fermi energy $\mu_{0}$ increases from $-3$.$26$, the lower band begins to be
occupied and $\mathcal{M}$ begins to decrease from $0$. When $\mu_{0}$
increases to $-1$.$74$, the lower band is fully occupied and the magnetization
decreases to $-0$.$7e/h$. When continue increasing $\mu_{0}$, the system
enters the first insulating region. The magnetization then linearly increases
with the Fermi energy until the middle band is occupied ($\mu_{0}$=$0^{-}$, at
this time $\mathcal{M}$=$1$.03$e/h$). When further increasing the Fermi
energy, the system immediately comes into the second insulating region and the
magnetization jumps down to $-1$.03$e/h$ and then linearly increases up to
$0.7e/h$ until the upper band begins to be occupied ($\mu_{0}$=$1$.$74$). Then
the magnetization decreases to $0$ with the Fermi energy going through the
upper band.

As shown in Refs. \cite{Wang2007, Wang2}, the totally different behavior of
the magnetization in the metallic and insulating regions is due to the
different roles $M_{c}$ and $M_{\Omega}$ play in these two regions. For
further illustration, we show in Fig. \ref{f3}(c) $M_{c}$ (dashed line) and
$M_{\Omega}$ (dotted line) as functions of the Fermi energy, their sum gives
$\mathcal{M}$ in Fig. \ref{f3}(b). One can see that overall $M_{c}$ and
$M_{\Omega}$ have opposite contributions to $\mathcal{M}$, which implies that
these two parts carry opposite-circulating currents. In each insulating regime
the conventional term $M_{c}$ keeps a constant, which is due to the fact that
the upper limit of the $k$-integral of $m_{n}(\mathbf{k})$ is invariant as the
chemical potential varies in the gap. In the metallic region, however, since
the occupied states varies with the Fermi energy, thus $M_{c}$ also varies
with $\mu_{0}$, resulting in a decreasing slope shown in Fig. \ref{f3}(c). The
Berry phase term $M_{\Omega}$ also displays different behavior between
insulating and metallic regions. In the insulating region, $M_{\Omega}$
linearly increases with $\mu_{0}$, as is expected from Eq. (\ref{m0}). The
slope of the Berry-phase correction term in insulating regions is proportional
to the system's Hall conductance. In the metallic region, however, this term
sensitively depends on the topological property of the band in which the
chemical potential is located. On the whole the comparison between Fig.
\ref{f3}(b) and Fig. \ref{f3}(c) shows that the metallic behavior of
$\mathcal{M}$ is dominated by its conventional term $M_{c}$, while in the
insulating regime $M_{\Omega}$ plays a main role in determining the behavior
of $\mathcal{M}$.\begin{figure}[ptb]
\begin{center}
\includegraphics[width=0.4\linewidth]{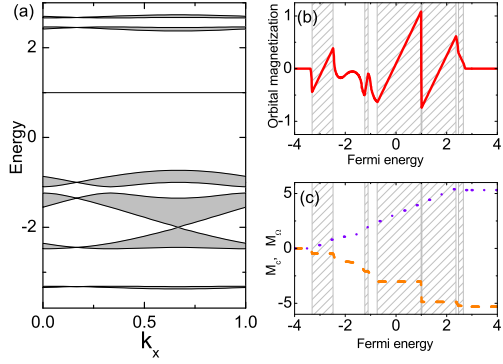}
\end{center}
\caption{(Color online). Energy spectrum and orbital magnetization for
$f$=$1/3$.}%
\label{f4}%
\end{figure}\begin{figure}[ptbptb]
\begin{center}
\includegraphics[width=0.4\linewidth]{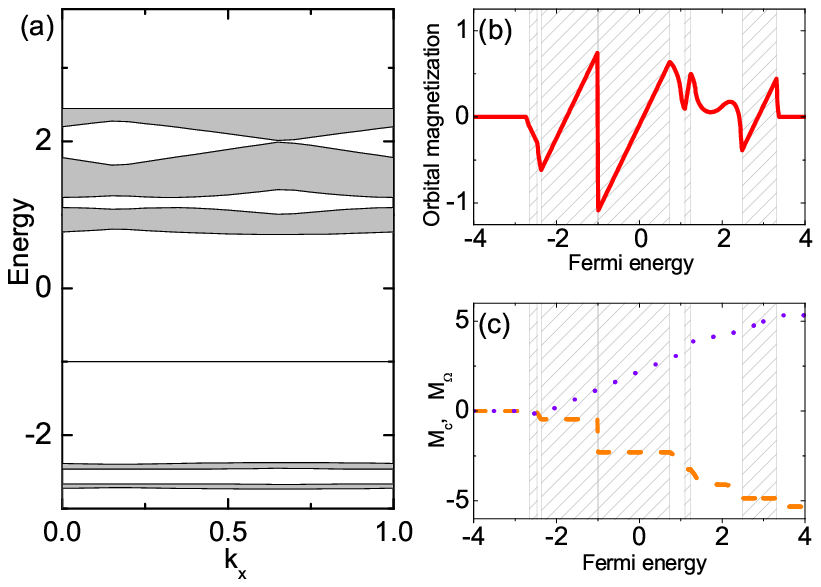}
\end{center}
\caption{(Color online). Energy spectrum and orbital magnetization for
$f$=$1/6$.}%
\label{f51}%
\end{figure}\begin{figure}[ptbptbptb]
\begin{center}
\includegraphics[width=0.4\linewidth]{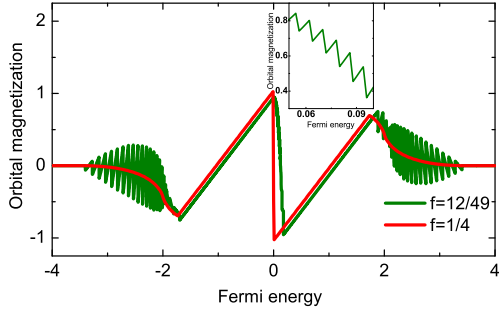}
\end{center}
\caption{(Color online). The magnetization $\mathcal{M}$ as a function of the
Fermi energy $\mu_{0}$ in the fields $f$=$1/4$ and $f^{\prime}$=$12/49$. The
inset is the enlarged vision of the magnetization $\mathcal{M}$ at $f^{\prime
}$=$12/49$ when the Fermi energy varies between $0.05$ and $0.1$.}%
\label{f6}%
\end{figure}\begin{figure}[ptbptbptbptb]
\begin{center}
\includegraphics[width=0.4\linewidth]{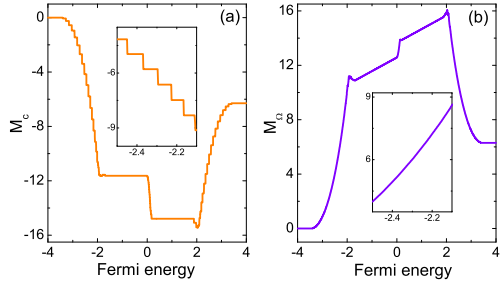}
\end{center}
\caption{(Color online). The two components of the magnetization $\mathcal{M}$
as functions of the Fermi energy $\mu_{0}$ at $f$=$12/49$.}%
\label{f8}%
\end{figure}

Now let us consider more complex cases, for example, with a field $f$=$1/3$
and with a field $f$=$1/6$. In these two case, there are more energy bands and
gaps appearing in the energy spectrum [see Figs. \ref{f4}(a) and
\ref{f51}(a)]. Figures \ref{f4}(b) and \ref{f4}(c), and Figs. \ref{f51}(b) and
\ref{f51}(c), for $f$=$1/3$ and $f$=$1/6$, respectively, plot the
magnetization $\mathcal{M}$ and its two components as functions of $\mu_{0}$.
Clearly from these figures, one can observe a similar variation of the
magnetization by changing the electron's fillings (the Fermi energy $\mu_{0}%
$). Comparing the case of $f$=$1/3$ [Fig. \ref{f4}(a)] to the case of
$f$=$1/6$ [Fig. \ref{f51}(a)], one can find that the orbital magnetization in
these two fields have a relation
\begin{equation}
\mathcal{M}(f=\frac{1}{6},\mu_{0})=\mathcal{M}(f=\frac{1}{3},-\mu_{0}).
\label{aaa}%
\end{equation}
This conclusion is a combined exhibition of the above mentioned symmetries
(ii) and (iii) in the Hofstadter energy spectrum.

We have known that the topology of the energy spectrum of the 2D lattice
system arises from the competition between the periodic potential-induced
level broadening and the magnetic field-induced level discretization. Once the
external magnetic field changes, the topology of the energy spectrum also
changes, which results in the simultaneous prominent variation in the
magnetization of the 2D system. The above discussions on the magnetization for
different fields $f$=$1/4$, $1/3$ and $1/6$ are typical examples. Now a
question is put forward: If there is a very small perturbation $\delta f$ to
the field $f$, then how about the change in the magnetization? The answer is
the occurrence of fractal structure in the magnetic de Haas-van Alphen (dHvA)
oscillations, which is very important for the experimental measurement of the
Fermi surface topology.

Now we investigate this fractal structure in the magnetic oscillations for the
present system. For convenience and simplicity, we choose the field $f$=$1/4$
and the perturbation $\delta f$=$-1/196$ ($f^{\prime}$=$f+\delta f$=$12/49$).
The calculated magnetization $\mathcal{M}$ at the fields $f$ and $f^{\prime}$
are plotted in Fig. \ref{f6}. From Fig. \ref{f6} one can observe the following
features: (i) In the insulating region, the magnitudes of the magnetization
$\mathcal{M}$ have little difference at the fields\ $f$ and $f^{\prime}$. The
perturbation $\delta f$ is more small, more little in the changes of the
magnetization $\mathcal{M}$; (ii) In the metallic regions at the field $f$,
however, there are dHvA oscillations appearing in the magnetization
$\mathcal{M}$ at the field $f+\delta f$. The perturbation $\delta f$ is more
small, the magnetization oscillates more rapidly; (iii) The collapsed middle
band originally at field $f$=$1/4$ now spreads at $f^{\prime}$=$12/49$. And in
this middle band there are also dHvA oscillations appearing in the orbital
magnetization. To see this fractal feature more clearly, we enlarge the
magnetization in the inset in Fig. \ref{f6}; (iv) There are similar variations
in the two components of $\mathcal{M}$ (see Fig. \ref{f8}). In the insulating
regions at $f$=$1/4$, there are little difference in the conventional and
Berry-phase terms. However, both terms exhibit different behaviors in the
metallic regions. While the Berry-phase term has little difference at
different fields as it in the insulating regions, there are many quantum steps
appearing in the conventional term at $f^{\prime}$=$12/49$. The reason for
these quantum steps appearing in the conventional magnetization is that the
topology of the energy spectrum at the field $f^{\prime}$=$f+\delta f$ is
different from that at the field $f$. When the external field $f$ changes to
$f+\delta f$, the eigenbands at the field $f$ split into lots of subbands. The
perturbation $\delta f$ is more small, the number of the splitting subbands is
more large. Because the upper limit of the $k$-integral of $m_{n}(\mathbf{k})$
is invariant as the Fermi level $\mu_{0}$ varies in the subgap, then the
conventional term in the subgap keeps a constant, which is the reason for the
quantum step appearing in the conventional term. On the other hand, the
integral of the Berry-phase term intimately depends on the Fermi level and has
no such relation with the energy gaps (or bands). That results in the little
difference in the Berry-phase term at different fields. So, when one changes
the electron's fillings through these splitting bands, the total magnetization
then exhibits the fractal structure in the magnetic dHvA oscillations.

\begin{figure}[ptb]
\begin{center}
\includegraphics[width=0.4\linewidth]{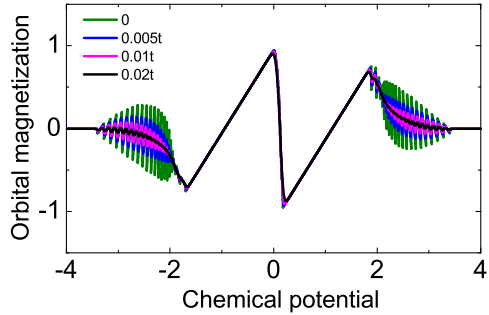}
\end{center}
\caption{(Color online). The magnetization $\mathcal{M}$ as a function of the
electron chemical potential at the field $f$=$12/49$ with different
temperatures.}%
\label{f9}%
\end{figure}

In the above discussions on the orbital magnetization of the 2D kagom\'{e}
lattice, we have concentrated on the zero-temperature limit and omitted
finite-temperature effect. Now let us briefly consider the more realistic
cases in which the finite-temperature effect is included. Figure \ref{f9}
plots the orbital magnetization at the field $f$=$12/49$ with different
temperatures $k_{B}T$=$0$, $0.005$, $0.01$, and $0.02$, respectively. From
Fig. \ref{f9}, one can find that the magnetic oscillations are suppressed by
thermal broadening. 

\section{Summary}

In summary, we have theoretically investigated the orbital
magnetization of a 2D kagom\'{e} lattice in a perpendicular magnetic
field. Here, the orbital magnetization includes a conventional term
and a Berry-phase term, which play different roles in metallic and
insulating regions. As examples, we have carefully discussed the
orbital magnetization and its two components at the fields
$f$=$1/4$, $1/3$, and $1/6$, respectively. By varying the Fermi
energy $\mu_{0}$, we have obtained the following results: (i) The
conventional term and the Berry-phase term give the opposite
contributions, with their magnitudes increasing (decreasing) with
increasing (decreasing) the Fermi energy in metallic regions; (ii)
The conventional term keeps unchanged in insulating regions; (iii)
The slope of the Berry-phase term in insulating regions is
proportional to the system's Hall conductance. When the flux is
applied near a commensurate one (for example, $1/4$), the magnetic
dHvA oscillations develop a fractal structure, i.e., the orbital
magnetization rapidly oscillates when the Fermi energy varies
through the split subbands. The finite-temperature effect has also
been shown to suppress the oscillating amplitude of the orbital
magnetization.

\begin{acknowledgments}
This work was supported by NSFC under Grants No. 90921003, No.
10904005, No. 60776061 and No. 60776063, and by the National Basic
Research Program of China (973 Program) under Grants No.
2009CB929103 and No. 2009CB929300. \qquad
\end{acknowledgments}


\begin{thebibliography}{99}                                                                                               %


\bibitem {Hof}D. R. Hofstadter, Phys. Rev. B \textbf{14}, 2239 (1976); G. H.
Wannier, Phys. Status Solidi B \textbf{88}, 757 (1978); G. H. Wannier, G. M.
Obermair, and R. Ray, Phys. Status Solidi B \textbf{93}, 337 (1979).

\bibitem {Claro}F. H. Claro and G. H. Wannier, Phys. Rev. B \textbf{19}, 6068 (1979).

\bibitem {Ram}R. Rammal, J. Phys. (Paris) \textbf{46}, 1345 (1985).

\bibitem {YXiao}Y. Xiao, V. Pelletier, P. M. Chaikin, and D. A. Huse, Phys.
Rev. B \textbf{67}, 104505 (2003).

\bibitem {Schu}M. A. Andrade Neto and P. A. Schulz, Phys. Rev. B \textbf{52},
14093 (1995).

\bibitem {Ando}H. Aoki, M. Ando, and H. Matsumura, Phys. Rev. B \textbf{54},
R17296 (1996).

\bibitem {Behrooz}A. Behrooz, M. J. Burns, H. Deckman, D. Levine, B.
Whitehead, and P. M. Chaikin, Phys. Rev. Lett. \textbf{57}, 368 (1986); K.
Springer and D. van Harlingen, Phys. Rev. B \textbf{36}, 7273 (1987); H.
Schwabe, G. Kasner, and H. B\"{o}ttger, Phys. Rev. B \textbf{56}, 8026 (1997).

\bibitem {Gordon}B. Douc\c{o}t, W. Wang, B. Pannetier, P. Rammal, A. Vareille,
and D. Henry, Phys. Rev. Lett. \textbf{57}, 1235 (1986); J. M. Gordon, A. M.
Goldman, J. Maps, D. Costello, R. Tiberio, and B. Whitehead, Phys. Rev. Lett.
\textbf{56}, 2280 (1986); Q. Niu and F. Nori, Phys. Rev. B \textbf{39}, 2134 (1989).

\bibitem {Vidal}J. Vidal, R. Mosseri, and B. Douc\c{o}t, Phys. Rev. Lett.
\textbf{81}, 5888 (1998).

\bibitem {Avron}O. Gat and J. E. Avron, Phys. Rev. Lett. \textbf{91}, 186801 (2003).

\bibitem {Avron1}O. Gat and J. E. Avron, New J. Phys. \textbf{5}, 441 (2003).

\bibitem {Mohan2003}P. Mohan, F. Nakajima, M. Akabori, J. Motohisa, and T.
Fukui, Appl. Phys. Lett. \textbf{83}, 689 (2003); P. Mohan, J. Motohisa, and
T. Fukui, Appl. Phys. Lett. \textbf{84}, 2664 (2004).

\bibitem {Hig2000}M. J. Higgins, Y. Xiao, S. Bhattacharya, P. M. Chaikin, S.
Sethuraman, R. Bojko, and D. Spencer, Phys. Rev. B \textbf{61}, R894 (2000);
Y. Xiao, D. A. Huse, P. M. Chaikin, M. J. Higgins, S. Bhattacharya, and D.
Spencer, \textit{ibid}. \textbf{65}, 214503 (2002).

\bibitem {Tong1985}S. Y. Tong, G. Xu, W. Y. Hu, and M. W. Puga, J. Vac. Sci.
Technol. B \textbf{3}, 1076 (1985).

\bibitem {Shi2001}K. Shiraishi, H. Tamura, and H. Takayanagi, Appl. Phys.
Lett. \textbf{78}, 3702 (2001).

\bibitem {Xiao}D. Xiao, J. Shi, and Q. Niu, Phys. Rev. Lett. \textbf{95},
137204 (2005).

\bibitem {Wang2007}Z. Wang and P. Zhang, Phys. Rev. B \textbf{76}, 064406 (2007).

\bibitem {Wang2}Z. Wang, P. Zhang, and J. Shi, Phys. Rev. B \textbf{76},
094406 (2007).

\bibitem {Chang}M.-C. Chang and Q. Niu, Phys. Rev. B \textbf{53}, 7010 (1996).

\bibitem {Sundaram}G. Sundaram and Q. Niu, Phys. Rev. B \textbf{59}, 14915 (1999).

\bibitem {Xiao2}D. Xiao, Y. Yao, Z. Fang, and Q. Niu, Phys. Rev. Lett.
\textbf{97}, 026603 (2006).

\bibitem {Thon1}T. Thonhauser, D. Ceresoli, D. Vanderbilt, and R. Resta, Phys.
Rev. Lett. \textbf{95}, 137205 (2005).
\end{thebibliography}
\end{document}